\documentclass[12pt]{iopart}
\usepackage[dvips]{graphicx}
\usepackage{latexsym}
\usepackage{graphicx}
\usepackage{dcolumn}
\usepackage{bm}
\usepackage{graphics}
\usepackage{epsfig,color}

\newcommand{\be}{\begin{equation}}
\newcommand{\ee}{\end{equation}}
\newcommand{\bea}{\begin{eqnarray}}
\newcommand{\eea}{\end{eqnarray}}

\begin{document}
\title{Some Aspects of Supersymmetric Field Theories with Minimal Length and Maximal Momentum}
\author{Kourosh Nozari$^{a}$\footnote{knozari@umz.ac.ir},\, F. Moafi$^{b}$  and  F. Rezaee Balef$^{b}$}\vspace{0.5cm}

\address{ $^{a}$Department of Physics, Faculty of Basic
Sciences,\\
University of Mazandaran, P. O. Box 47416-1467, Babolsar, IRAN}
\vspace{0.25cm}
\address{ $^{b}$Department of Physics, Islamic Azad University, Sari Branch, Sari, IRAN}
\date{\today}
\begin{abstract}
We consider a real scalar field and a Majorana fermion field to
construct a supersymmetric quantum theory of free fermion fields
based on the deformed Heisenberg algebra $[x,p]=i\hbar\big(1-\beta
p+2\beta^{2}p^{2}\big)$, where $\beta $ is a deformation parameter.
We present a deformed supersymmetric algebra in the presence of
minimal length and maximal momentum.\\

\end{abstract}

\maketitle

\section{Introduction}\label{sec1}
Physics in extremely high energy regions is particularly of interest
to particle physicists. One of the promising idea in the realm of
high energy physics is the idea of supersymmetry (SUSY) (see for
instance [1] and references therein). This is a symmetry that
relates or mixes (unites) fermions and bosons. Supersymmetry has the
potential to solve some outstanding mysteries in particle physics.
One important problem that stands out is called the \emph{hierarchy
problem}. It is believed that the mass of the Higgs boson, $m_{h}$,
is much smaller than the fundamental Planck mass, $m_p
=\sqrt{\frac{\hbar c}{G}}\approx 10^{19}$ GeV/$c^{2}$. On the other
hand, when we discuss gravity in the context of the quantum field
theory, it is expected that there is a minimal measurable length
that restricts resolution of adjacent spacetime points up to the
Planck length. This is actually a common address of all quantum
gravity candidates (see [2] and references therein). Incorporation
of this idea in quantum field theory provides a theory that is
naturally regularized in the ultra-violate regime [3,4,5]. Recently
in the context of the Doubly Special Relativity it has been revealed
that a test particle's momentum cannot be arbitrarily imprecise
leading nontrivially to the result that there is a maximal momentum
for test particles [6]. This idea can be explained also through
modified dispersion relations as a phenomenological outcome of
foam-like structure of spacetime at the Planck scale [7,8]. String
theory which has a characteristic scale $\sqrt{\alpha}$, is one of
the most successful theoretical framework which overcomes the
difficulty of ultra-violet divergence in quantum theory of gravity.
Therefore, if we construct a field theory which captures some
stringy nature and/or includes stringy corrections, it would play a
pivotal role in investigating physics in high energy regions even
near the Planck scale. One way to discuss these corrections is
deforming the standard Heisenberg uncertainty principle to a
generalized uncertainty principle (GUP). A GUP that predicts a
minimal observable length can be written as follows

\begin{equation}
\Delta x\Delta p\geq\frac{\hbar}{2}+{\beta}_{0} \ell_{pl}^2
\frac{(\Delta p)^2}{\hbar},
\end{equation}
This GUP leads to the following algebraic structure

\begin{equation}
[x,p]=i\hbar(1+\beta^{2}p^{2})
\end{equation}
where $\beta={\beta}_{0} /({M}_{pl} c)^2  = {\beta}_{0} {l^2}_{pl}
/\hbar^2 $. If GUP is realized in a certain string theory context,
$\beta$ takes a value of the order of the string scale, $\beta \sim
\alpha$. In the presence of both minimal length and maximal
momentum, the GUP can be given as follows
\begin{eqnarray}
{\Delta x\Delta p\geq\frac{\hbar}{2}\big(1-2\beta\langle
p\rangle+4\beta^2\langle p^2\rangle\big)}.
\end{eqnarray}
In this framework the following algebraic structure can be deduced
[9]
\begin{eqnarray}
[x,p]=i\hbar(1-\beta p+2\beta^2 p^2)\,.
\end{eqnarray}
In this context, we can define the generalized momentum operator as
\begin{eqnarray}
{\textbf{P}=p(1-\beta p+2\beta^2 p^2)}
\end{eqnarray}
With these preliminaries, in this paper we construct a quantum
theory of free fermion fields based on the deformed Heisenberg
algebra (4). We respect the supersymmetry in this context as a
guiding principle. This is because string theory contains this
symmetry and it provides a useful tool to understand physics in
ultra-violet regime. Because of existence of natural cutoffs as
minimal length and maximal momentum, the supersymmetry is deformed
in this context. From the fermionic part of the system, we propose
an action of fermionic field based on the GUP in a spacetime with
arbitrary number of dimensions. We construct the deformed
supersymmetric algebra in this context. Finally, we compute some
scattering amplitudes such as the pair annihilation amplitude to see
how these important quantities can be calculated in the presence of
natural cutoffs. We note that experimental limits and the general
phenomenology perspective on this issue are not discussed here. One
can see Ref. [10] for this purpose. We note also that while the
study of SUSY in relation to GUP is a relatively young subject, a
bit more has been done for other quantum-spacetime intuitions. For
instance, in Ref. [11] the authors have studied the basic twist
quantization of $osp(1|2)$ and kappa-deformation of $D=1$
superconformal mechanics.

The organization of the paper is as follows: in Section 2 we
construct a scalar field theory based on the GUP (4). Section 3 is
devoted to the issue of supersymmetry in the presence of natural
cutoffs as minimal length and maximal momentum. In Section 4 we
compute scattering amplitude of the pair annihilation in the
presence of natural cutoffs. The paper follows by a summary and
conclusion in section 5.

\section{Scalar Field Theory}\label{sec2}

Generalization of the Heisenberg algebra to $d$-dimensions, where
rotational symmetry is preserved and there are both a minimal length
and maximal momentum as natural cutoffs, is as follows
\begin{eqnarray}
{\textbf{[}\textbf{x}_i,\textbf{p}_j\textbf{]}=i\hbar
(1-\beta \textbf{p}+2\beta^2 \textbf{p}^2)\delta_{ij} },
\end{eqnarray}

Here $i$ and $j$ run from $1$ to $d$, where in three dimensions
$\textbf{p}=p_x\textbf{i}+p_y\textbf{j}+p_z\textbf{k}$ that $
\textbf{i}$, $\textbf{j}$ and $\textbf{k}$ are unit vectors of
cartesian coordinates and $ \textbf{p}^2=\sum_{i=1}^{d}(p_i)^2 $ .
Hereafter, we use indices $i$ and $j$ for spatial coordinates and
$a$ and $b$ for all spacetime coordinates. Now the Jacobi identity
determines the full algebra of the theory which is given as follows
(see the work by K. Nozari and A. Etemadi in Ref. [6])

\begin{eqnarray}
{\textbf{[}\textbf{x}_i,\textbf{p}_j\textbf{]}=i\hbar (1-\beta
\textbf{p}+2\beta^2 \textbf{p}^2)\delta_{ij} },
\end{eqnarray}
\begin{eqnarray}
{[\textbf{p}_i , \textbf{p}_j ]=0},
\end{eqnarray}
\begin{eqnarray}
[\textbf{x}_i,\textbf{x}_j] = -i \beta \hbar\ (4 \beta -
\frac{1}{\textbf{p}})\ (1 - \beta \textbf{p} + 2\beta^2
\textbf{p}^2)\ \textbf{L}_{ij}\,.
\end{eqnarray}

The presence of the $\frac{1}{\textbf{p}}$ is a trace of the
existence of the maximal momentum in this setup. Note also that Eq.
(9) reflects the noncommutative nature of the spacetime manifold in
Planck scale. Here $\textbf{L}_{ij} $ are angular momentum like
operators defined as

\begin{eqnarray}
{\textbf{L}_{ij}=\frac{1}{1-\beta \textbf{p}+2\beta^2 \textbf{p}^2}
\big(\textbf{x}_i \textbf{p}_j -\textbf{x}_j \textbf{p}_i) },
\end{eqnarray}

Since operators $\textbf{p}^i$'s commute with each other and we have
not assumed the existence of a minimal uncertainty in momentum (see
Ref. [12] for possible extension in this regard), we can construct
our theory in momentum space representation. In momentum space
representation, momentum operators are diagonalized simultaneously
and we do not distinguish eigenvalues of momentum $\textbf{p}_i$
from operator $ \textbf{p}^i$. In which follows, we set $ \hbar=1 $
for simplicity. The Lagrangian in $ d+1 $ dimensional spacetime and
in the presence of a minimal observable length and a maximal
momentum can be written as follows
\begin{eqnarray}
{\mathcal{L}=-\frac{1}{2}\int_{-p_{pl}}^{+p_{pl}} d^{d}p
\Big(1-\beta \textbf{p}+2\beta^2 \textbf{p}^2\Big)^{-1}
\Phi(-\textbf{p},t)
\Big[{\partial}_{t}^2+\textbf{p}^2+m^2\Big]\Phi(\textbf{p},t)}
\end{eqnarray}

where
\begin{eqnarray}
{\textbf{p}^2=\sum_{i=1}^{d}(p_i)^2 }\nonumber\,.
\end{eqnarray}
Note that the difference from ordinary quantum field theory is the
extra prefactor $(1-\beta \textbf{p}+2\beta^2 \textbf{p}^2)^{-1}$ in
the Lagrangian. Using the Bjorken-Johnson-Low prescription, from
behavior of $T^{*}$-product through the following relations
\begin{eqnarray}
{\lim
_{\textbf{q}\longrightarrow\infty}<T^{\ast}\hat{\Phi}(\textbf{p},\textbf{q})\hat{\Phi}(\textbf{p}',\textbf{q}')>}
\nonumber
\end{eqnarray}

\begin{eqnarray}
{\lim _{\textbf{q}\longrightarrow\infty} \textbf{q}
<T^{\ast}\hat\Phi(\textbf{p},\textbf{q})
\hat{\Phi}(\textbf{p}',\textbf{q}')>}
\end{eqnarray}
where
\begin{eqnarray}
{T^{\ast}\hat{\Phi}(\textbf{p},\textbf{q)}\hat{\Phi}(\textbf{p}',\textbf{q}')
= T\hat{\Phi}
(\textbf{p},\textbf{q})\hat{\Phi}(\textbf{p}',\textbf{q}')},\nonumber
\end{eqnarray}
we obtain
\begin{eqnarray}
{\Big[ \hat{\Phi}(\textbf{p},t) , \dot{\hat{\Phi}}(\textbf{p}',t)
\Big] = i\hbar(1-\beta \textbf{p}+2\beta^2 \textbf{p}^2)
\delta(\textbf{p}+\textbf{p}')}
\end{eqnarray}
where
\begin{eqnarray}
{ \hat{\Phi}(\textbf{p},t)=\frac{\hbar}{E(\textbf{p})}
\hat{\phi}(\textbf{p}) \exp\Big(\frac{t}{i\hbar}E(\textbf{p})\Big)+
\frac{\hbar}{E(-\textbf{p})} \hat{\phi^\dag}(-\textbf{p})}
{}\nonumber {\exp\Big(\frac{-t}{i\hbar}E(-\textbf{p})\Big)}.
\end{eqnarray}
In this relation,  $E(\textbf{p})\equiv\sqrt{Z(\textbf{p})+m^{2}}$
where $Z(\textbf{p})$ is an arbitrary even function whose explicit
form has no influence on the arguments. One can see from equation
(13) that a deformation prefactor, $(1-\beta \textbf{p}+2\beta^2
\textbf{p}^2) $, of Heisenberg algebra in the first quantization (6)
also appears in canonical commutation relation of the second
quantized field theory.

In the fermion field case, we encounter a difficulty in the
construction of the second quantized Hilbert space which dose not
appear in a scalar field system. In which follows we use the idea of
supersymmetry to construct a quantized field theory of fermions. In
fact, we construct a quantum Field Theory of fermions, which is
consistent with the above scalar field theory, and by using
supersymmetry prescription with a minimal length and maximal
momentum scale.


\section{Supersymmetry and GUP} \label{sec3}

Supersymmetric quantum field theory with minimal length has been
studied by Shibusa [13]. In which follows, we generalize the pioneer
work of Shibusa to the more general case that there are both minimal
length and maximal momentum as natural cutoffs.

As we have mentioned in the introduction, supersymmetry proposes
that to each fermion there exists a boson and vice versa. Thus in
two and three-dimensional spacetime, a system with a real scalar and
a Majorana fermion has a special symmetry between a boson and a
fermion, namely SUSY. Following Ref. [13], our notation for two and
three-dimensional spacetime is as follows: in those dimensional
spacetime (with signature $(- , +)$  or $(- , +, +)$) the Lorantz
group has a real (Majorana) two-component spinor representation $
\psi^\alpha $.  For instance, in three-dimensional spacetime, we
define a representation of Gamma matrices by Pauli matrices as
follows [13]:

\begin{eqnarray}
{\{\Gamma^a , \Gamma^b \}=2\eta^{ab}=2\, diag(- + + )}
\end{eqnarray}

\begin{eqnarray}
{\Gamma^0 =-i\sigma_2=i\sigma^2,\quad\quad
\Gamma^1=\sigma_1=-\sigma^1, \quad\quad  \Gamma^2=-\sigma_3} .
\end{eqnarray}
We note that the spinor indices in this case are lowered or raised
by charge conjugation matrix $C_{\alpha\beta}\equiv\Gamma^0$ and its
inverse matrix $C^{-1}$. The generalized supersymmetric algebra in
the presence of a minimal length and a maximal momentum and its
action on a scalar field $\phi$, a Majorana fermion $\psi$ and an
auxiliary field $F$ with parameter $\epsilon^\alpha $ is as follows
\begin{eqnarray}
[\bar{\epsilon}_1Q ,
\bar{\epsilon}_2Q]=2\Upsilon\bar{\epsilon}_1\Gamma^a\epsilon_2P_a,
\end{eqnarray}
\begin{eqnarray}
{\delta\phi(p,t)=i\bar{\epsilon}\psi(p,t)},
\end{eqnarray}
\begin{eqnarray}
{\delta\psi^\alpha(p,t)=\lambda_{1} F(p,t)\epsilon^\alpha-\lambda_2\{(\bar{\epsilon}
\Gamma^0C^{-1})^\alpha\partial_t+(\bar{\epsilon}\Gamma^jC^{-1})^\alpha
(ip_j)\}\phi(p,t),}
\end{eqnarray}
\begin{eqnarray}
{\delta
F(p,t)=\lambda_3i\bar{\epsilon}[\Gamma^0\partial_t+\Gamma^{j}(ip_j)]\psi(p,t)},
\end{eqnarray}
where $\Upsilon$ and $\lambda_i$ are functions of GUP deformation
parameter $\beta$ and momentum. These factors reduce to unity in the
limit of $\beta\rightarrow0$ and will be determined later by
consistency condition. The closeness of algebra requires
\begin{eqnarray}
{\lambda_1\lambda_3=\lambda_2=\Upsilon.}
\end{eqnarray}
Now the supersymmetric Lagrangian can be written as a sum of
separate Lagrangians, that is,
\begin{eqnarray}
{\cal{L}}={\cal{L}}_B+{\cal{L}}_F+{\cal{L}}_{aux}\nonumber,
\end{eqnarray}
where ${\cal{L}}_B$, ${\cal{L}}_F$ and ${\cal{L}}_{aux}$ are
Bosonic, Fermionic and auxiliary fields Lagrangian respectively.
Following [13], we generalize the Lagrangian by introduction of
factors $\zeta_i$, which are functions of  deforming parameter
$\beta$ and momentum. These functions can be determined as follows.
Starting with Lagrangian
\begin{eqnarray}
{\mathcal{L}=-\frac{1}{2}\int_{-p_{pl}}^{+p_{pl}} dp^d\Big\{\zeta_1\phi(-p,t)
({\partial_t}^2+\textbf{p}^2+m^2)\phi(p,t)}\nonumber\\
{}\nonumber\\
{+i\zeta_2\bar{\psi}(-p,t)
(\Gamma^0\partial_t+(ip_i)\Gamma^i+m)\psi(p,t)}\nonumber\\
{}\nonumber\\
{+\sqrt{\zeta_1\zeta_3}
m\phi(-p,t)F(p,t)-\zeta_3F(-p,t)F(p,t)\Big\}},
{}\nonumber\\
\end{eqnarray}
where $d$ is the number of the spatial coordinates. By integrating
out the field $F$, we obtain the Lagrangian for the scalar and
Majorana fields as follows
\begin{eqnarray}
{\mathcal{L}=-\frac{1}{2}\int_{-p_{pl}}^{+p_{pl}}
dp^d\Big\{-\zeta_1\phi(-p,t)
({\partial_t}^2+\textbf{p}^2+m^2)\phi(p,t)}\nonumber\\
{}\nonumber\\
{+i\zeta_2\bar{\psi}(-p,t)
(\Gamma^0\partial_t+(ip_i)\Gamma^i+m)\psi(p,t)\Big\}.}
\end{eqnarray}
On the other hand, the invariance of the Lagrangian (21) under
supersymmetry variation (17)-(19) in the presence of the minimal
length and maximal momentum leads to
\begin{eqnarray}
{\lambda_1\zeta_2=\sqrt{\zeta_1\zeta_3},}\nonumber
\end{eqnarray}
\begin{eqnarray}
{\lambda_1\zeta_2=\lambda_3\zeta_3,}\nonumber
\end{eqnarray}
\begin{eqnarray}
{\zeta_1=\lambda_1\lambda_3\zeta_2.}
\end{eqnarray}
Using the conditions (20) and (23), only $\zeta_{1}$ and $\zeta_{2}$
need to be determined. Note also that factor $\lambda_1$ can be
absorbed into normalization of $F$. In which follows we set
$\lambda_1=1$ for $F$ to be an auxiliary field. Noether's current
for supersymmetric Lagrangian (22) can be calculated and the
supersymmetric charge is obtained to be
\begin{eqnarray}
{Q^\alpha=\int dt\int_{-p_{pl}}^{+p_{pl}} dp^d \zeta_1
\{-\psi^\alpha(-p,t)
\partial_t\phi(p,t)+(\Gamma^i\Gamma^0\psi(-p,t))^\alpha}\nonumber\\
{}\nonumber\\
{\times(ip_i)\phi(p,t)+m(\Gamma^0\psi(-p,t))^\alpha\phi(p,t)\}.}
\end{eqnarray}
Now the Hamiltonian of this system can be written as follows
\begin{eqnarray}
{\mathcal{H}=P^0=-\frac{1}{4}\frac{\zeta_2}{\zeta_1}(C\Gamma^0)_{\alpha\beta}
\{Q^\alpha , Q^\beta\}.}
\end{eqnarray}
Using the Bjorken-Johnson-Low prescription, from behaviors of
$T^\ast$-product between fields, as has been explained in the
previous section, we obtain canonical relations as follows
\begin{eqnarray}
{[\phi(p,t),\partial_t\phi(q,t)]=i{\zeta_1}^{-1}\delta(p+q),}
\end{eqnarray}
\begin{eqnarray}
{\{\psi^\alpha(p,t) , \psi^\beta(q,t)\}=-{\zeta_2}^{-1}
(\Gamma^0C^{-1})^{\alpha\beta}\delta(p+q).}
\end{eqnarray}
Therefore, the Hamiltonian (25) now can be written as follows
\begin{eqnarray}
{\mathcal{H}=\frac{1}{2}\int_{-p_{pl}}^{+p_{pl}} dp^d\zeta_1\{ \pi(-p,t)
\pi(p,t)+\phi(-p,t)(\textbf{p}^2+m^2)\phi(p,t)\}}\nonumber\\
{}\nonumber\\
{+i\zeta_2\bar{\psi}(-p,t)\Big((ip_i)\Gamma^i+m\Big)\psi(p,t),}
\nonumber\\
\end{eqnarray}
where $ \pi(p,t)\equiv\partial_t\phi(-p,t)$ and $i=1,2,...,d$. There
is another condition that can be used to determine $\zeta_1$ and
$\zeta_2$ and comes from the free energy of the supersymmetric
vacuum. From algebra (16), the supersymmetric state has zero energy
\begin{eqnarray}
{0=\frac{1}{2}{Tr}_B\ln \big[\zeta_1(E^2+\textbf{p}^2+m^2)\big]-
\frac{1}{4}{Tr}_F\ln \big[\zeta_1(E^2+\textbf{p}^2+m^2)\big],}
\end{eqnarray}
where $ {Tr}_B$ and ${Tr}_F$ represent trace in bosonic and
fermionic Hilbert space respectively. This relation leads to the
following condition
\begin{eqnarray}
{\zeta_1=\zeta_2^2.}
\end{eqnarray}
Finally we can write $\zeta_1= \Big(1-\beta \textbf{p}+
2\beta^2\textbf{p}^2\Big)^{-1}$ as one can read from (11). Thus we
have
\begin{eqnarray}
{\Upsilon=\lambda_2=\lambda_3=\zeta_2=\Big(1-\beta \textbf{p}+
2\beta^2\textbf{p}^2\Big) ^{-\frac{1}{2}},}
\end{eqnarray}
\begin{eqnarray}
{\lambda_1=\zeta_3=1,}
\end{eqnarray}
\begin{eqnarray}
{\zeta_1=\Big(1-\beta \textbf{p}+ 2\beta^2\textbf{p}^2\Big)^{-1}.}
\end{eqnarray}
Therefore, we have constructed the quantized field of fermions as
follows
\begin{eqnarray}
{\{\psi^\alpha(p,t) , \psi^\beta(q,t)\}=-\Big(1-\beta \textbf{p}+
2\beta^2\textbf{p}^2\Big)^{\frac{1}{2}}(\Gamma^0C^{-1})^{\alpha\beta}\delta(p+q).}
\end{eqnarray}
in consequence, the supersymmetric algebra in the presence of both
minimal length and maximal momentum is deformed from the usual one
as follows
\begin{eqnarray}
{[\bar{\epsilon}_1Q , \bar{\epsilon}_2Q]=2(1-\beta \textbf{p}+2\beta^2
 \textbf{p}^2)^{-\frac{1}{2}}\bar{\epsilon}_1\Gamma^a\epsilon_2p_a.}
\end{eqnarray}
These results can be generalized to higher dimensions easily.
Finally we note that the Lagrangian now takes the following form
\begin{eqnarray}
{\mathcal{L}=\int_{-p_{pl}}^{+p_{pl}} d^{d}p \Big\{- i\Big(1-\beta \textbf{p}+ 2\beta^2\textbf{p}^2\Big)^{-\frac{1}{2}}\bar{\psi}(-p,t) }\nonumber\\
{}\nonumber\\
{\times\big[\Gamma^0\partial_t+(ip_i)\Gamma^i+m\big]\psi(p,t)\Big
\}.}
\end{eqnarray}
We see that in the presence of quantum gravity effects (as minimal
length and maximal momentum) there is an extra, universal prefactor
$ \Big(1-\beta \textbf{p}+ 2\beta^2\textbf{p}^2\Big)^{-\frac{1}{2}}$
in comparison with the usual fermion action, regardless of existence
or absence of supersymmetry. We note that the major difference
between our framework and the formalism presented in Ref. [13] is
the difference in the mentioned prefactor. Our prefactor contains
both minimal length and maximal momentum simultaneously. Otherwise,
the algebraic structure of the two supersymmetric algebras are the
same.


\section{Some Scattering Amplitudes in the Presence of Minimal Length and Maximal Momentum }\label{sec4 }
In a perturbative expansion one usually refines the calculations
using corrections that are becoming smaller and smaller. The
reduction in importance is quantified by the power of the
perturbation parameter, which is usually the coupling strength. At
some point we are adding refinements that are too small to be
measured and we know we can stop adding refinements. In a
perturbation expansion, the amplitude $M$ for a given process can be
computed using an expansion of the type [14]
\begin{eqnarray}
{M=\sum_n g^{k_n}M_n}
\end{eqnarray}
where $g^{k_n}$ is the coupling constant. In which follows we are
going to see the effect of minimal length and maximal momentum as
natural cutoffs on the scattering amplitude of the
annihilation-creation process with spin-0 boson. In the presence of
just a minimal measurable length as ultra-violet cutoff, momentum
generalizes to $p\rightarrow p(1+\beta p^{2})$. So we have
\begin{eqnarray}
{M=\int_{-\infty}^{+\infty}\frac{-ig^2dp}{(1+\beta
p^2)\Big[p^{2}(1+\beta p^2)^2-m_{B}^{2}\Big]}}
\end{eqnarray}

We have discard the remaining delta function which is

\begin{eqnarray}
{(2\pi)^4\delta(p_1+p_2-p_3-p_4)},\nonumber
\end{eqnarray}
that $p_{i}$ are defined in the geometry of the process as shown in
figure 1.

\begin{figure}[htp]
\begin{center}
\includegraphics{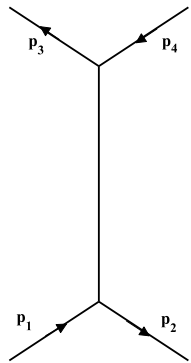}\vspace{6cm} \caption{\scriptsize{The geometry of the
annihilation-creation process.}}
\end{center}
\end{figure}

Simplifying relation (38), we find

\begin{eqnarray}
{M=\int_{-\infty}^{+\infty}\frac{-ig^2dp}{p(1+\beta
p^2)^3-{m_{B}(1+\beta p^2)^2}}}
\end{eqnarray}
By solving the integral, we obtain

\begin{eqnarray}
{M={-ig^2}\Bigg\{\frac{1}{{m_{B}}{\beta}}\Big[\frac{1}{(1+\beta p)}
-\frac{2\tanh^{-1}(\frac{2\beta
P+1}{\sqrt{4{m_{B}\beta}+1}})}{\sqrt{4{m_{B}\beta}+1}}\Big]}\nonumber\\
+\frac{1}{{m_{B}}\beta^{2}}\Big[{{\ln(1+\beta p)}}
-\frac{\tanh^{-1}(\frac{2\beta
p+1}{\sqrt{4{m_{B}\beta}+1}})}{\sqrt{4{m_{B}\beta}+1}}
-\frac{1}{2}{\ln(\beta^{2}P^{2}+\beta P-{m_{B}}\beta)}\Big]\Bigg\}
\end{eqnarray}
The probability for the process to occur is given by $|M|^{2}$.
Now we calculate the scattering amplitude in the presence of both
minimal length and maximal momentum. In this case, $p\rightarrow
p(1-\beta p +2\beta^{2} p^{2})$ and we have

\begin{eqnarray}
{M=\int_{-p_{pl}}^{+p_{pl}}-ig^2\frac{dp}{\Big[\Big(p(1-\beta
p+2\beta^2 p^2)\Big)^{2} -{m_{B}}^{2}\Big](1-\beta p+2\beta^2
p^2)}}
\end{eqnarray}
which can be simplified to find
\begin{eqnarray}
{M=\int_{-p_{pl}}^{+p_{pl}}-ig^2\frac{dp}{p(1-\beta p+2\beta^2 p^2)
^3-m_{B}(1-\beta p+2\beta^2 p^2)}}\,.
\end{eqnarray}
By solving this integral, we find the following result for
scattering amplitude in the presence of both minimal length and
maximal momentum

\begin{eqnarray}
{M={-ig^2}\frac{1}{2\beta^{2}m_{B}^{2}}\Bigg\{{\sum_{R}
\frac{(1+R)^{2}}{3R^{2}+2R+7}\ln(4\beta p_{pl}-1-R)}}\nonumber\\
{}\nonumber\\
-{\sum_{R} \frac{(1+R)^{2}}{3R^{2}+2R+7}
{\ln(-4\beta p_{pl}-1-R)}}\nonumber\\
{}\nonumber\\
+{\frac{1}{2}\ln\frac{(4\beta p_{pl}+1)^{2}+7}{(4\beta
p_{pl}-1)^{2}+7} -\frac{1}{\sqrt{7}}\Big({\tan^{-1}\frac{4\beta
p_{pl}-1}{\sqrt{7}}}
+{\tan^{-1}\frac{4\beta p_{pl}+1}{\sqrt{7}}}\Big)\Bigg\}}\nonumber\\
{}\nonumber\\
+{\frac{8ig^{2}}{7\beta m_{B}}\Bigg\{\frac{4\beta p_{pl}-1}{(4\beta
p_{pl}-1)^{2}+7}+ \frac{4\beta p_{pl}+1}{(4\beta p_{pl}+1)^{2}+7}+
{\frac{1}{\sqrt{7}}\Big({\tan^{-1}\frac{4\beta p_{pl}-1}{\sqrt{7}}})}}\nonumber\\
+{\tan^{-1}\frac{4\beta p_{pl}+1}{\sqrt{7}}}\Big)\Bigg\}\nonumber\\
{}\nonumber\\
\end{eqnarray}
where by definition $R$ is the root of
$2z^{2}+7z+7-32\beta^{32}m_{b}=0$. Now the probability for the
process to occur in the presence of these natural cutoffs is given
by $|M|^{2}$.


\section{Summary}\label{sec6 }

In this paper we have constructed a quantum theory of free fermion
fields based on the deformed Heisenberg algebra that contains both a
minimal measurable length and a maximal momentum for test particles.
Our strategy was to respect the supersymmetry in this context as a
guiding principle. This is because string theory contains this
symmetry and it provides a useful tool to understand physics in
ultra-violet regime. Due to existence of natural cutoffs as minimal
length and maximal momentum, the supersymmetry is deformed in this
context. From the fermionic part of the system, we proposed an
action of fermionic field based on the GUP in a spacetime with
arbitrary number of dimensions. We have constructed the deformed
supersymmetric algebra in this context. Finally, we have computed
the scattering amplitude for pair annihilation to see how these
important quantities can be calculated in the presence of natural
cutoffs as minimal length and maximal momentum.

\end{document}